\documentclass[useAMS,usenatbib]{mn2e}
\usepackage[dvips]{graphicx}

\title[Spatial variations in the field]{Spatial variations in the field of velocities and real solar granulation}
\author[M. I. Stodilka and S. Z. Malynych]{M. I. Stodilka\thanks{E-mail:
sun@astro.franko.lviv.ua} and S. Z.
Malynych\\
Astronomical Observatory, Ivan~Franko National University of Lviv, 8 Kyryla \& Methodia Str., Lviv 79005, Ukraine}
\begin{document}

\date{Accepted 2006 . Received 2006 ; in original form 2006 }

\pagerange{\pageref{firstpage}--\pageref{lastpage}} \pubyear{2006}

\maketitle

\label{firstpage}

\begin{abstract}
In this paper, the physical conditions within the inhomogeneous solar atmosphere have been reconstructed by means of solving the inverse problem of Non Local Thermodynamic Equilibrium (NLTE) radiative transfer. The profiles of $\lambda=523.42$ nm FeI spectral line of high spatial and time resolution were used as observational data. The velocity field has been studied for the real solar granulation in superadiabatic layer and overshooting convection region. Also, we investigate the vertical structure of inhomogeneous solar photosphere and consider penetration of granules from convective region into upper layers of stable atmosphere. The microturbulent velocity appears to be minimal at the bottom of overshooting convection region and increases sharply through superadiabatic layer and upper photosphere. High-turbulent layers emerge either in the central part of a flow or at the boundary of an incipient flow with following drift toward the centre of the flow. Wide descending flows tend to disintegrate into structures having  turbulence augmented, these structures correspond to the flows of matter. High microturbulence of the intensive flows provokes steep temperature depression in upper photosphere leading to the second inversion of temperature for the intergranules. The inversion of vertical velocities is observed to be frequent in the solar granulation. Some of the convective flows reach the minimum temperature region. Vertical convective velocities of the matter flows were found to be smaller in the middle and upper photosphere. Also, the effect of finite resolution on the spacial variations of the velocities in solar photosphere has been estimated.
\end{abstract}

\begin{keywords}
granulation -- photosphere: Sun.
\end{keywords}

\section{Introduction}

As it is known, energy transfer changes from entirely convective to radiative one on Sun's surface. The interaction between convection and radiation determines the  structure of the middle atmosphere, observed granulation features, and the solar activity as well. It is common to study the structure of solar atmosphere and convective motions applying both correlational and spectral analysis to the line profiles. The analysis of the radiation coming from Sun's surface provides an information about physical conditions of the plasma, i.e. the temperature, field of velocities, magnetic field, etc. In particular, spectral lines of heavy elements with Doppler widths less than those for typical photosphere velocities enable direct probing of the field of velocities. Also, the convection velocity, velocity of the undulating motion, micro- and macroturbulence velocity can be probed in this way along with their fluctuations.  The spectral lines of neutral Iron are especially valuable in this regard for solar (stellar) plasma probing, since they prevail in solar spectrum. Moreover, the atomic data for neutral Iron exceed in accuracy those for FeII.

The radiation hydrodynamic simulations of the granulation were carried out to investigate the field of velocities for solar and stellar atmospheres \citep{gadunploner, asplund, gadun, li, nordlund, ploner, robinson, spruit, stein}. To bring the results of simulations into accordance with observations the former results can be spatially smeared. Recently, an interest to the inverse methods of studying the atmosphere of the Sun and stars has been growing up \citep{ruiz, socas-navarro, socas, teplitskaya}. Those methods allow obtaining the information about physical processes affecting the radiation directly from observations.
The stratification of both temperature and velocity field in the solar granulation was studied by \citet{frutiger, borrero}, although these papers addressed mostly the temperature granulation structure. The profiles of high spectral, but low spatial resolution have been used therein. Thus, the granulation models obtained by inverse technique represent some averaged properties. \citet{frutiger} applied the regularization to the solution of the inverse code to damp oscillations in the derivable stratifications of reproducible parameters of the models. In both cited papers NLTE effects were not taken into account, while microturbulent velocity considered to be constant (i.e. not varying with the altitude). The approach proposed by \citet{frutiger} does not take into account the granulation structure at small scales and individual properties of the convective cells as well.

Because the solutions of the inverse problem tend to have spurious oscillations the applications of high spatial resolution profiles to the solar granulation studies are quite rare \citep{rodriguez}. The analysis of global spatial variations of the parameters of the spectral lines of neutral Iron with high spatial resolution has been done by \citet{puschmann}. Also, \citet{PRVBH05} have studied space-time distribution of thermodynamic values and vertical velocities of flows by means of inversion method basing on neutral Iron spectral lines with high spatial and time resolution. The penetration of granules into photosphere has been studied; the inversion of temperature and density of matter has been revealed. However, for some reason the authors used NLTE parameters for undisturbed atmosphere, although there is a number of NLTE inverse codes. This is not correct, if it is going about probing of physical conditions in upper photosphere layers. 

According to \citet{esp} the photosphere can be divided in two components with different physical conditions having the interface situated at $\sim 170$ km above the continuum formation level. 

The granulation brightness and convective velocities in the solar photosphere between the levels of the formation of continuum radiation and the temperature minimum were examined in \citep{kshch}.

The spectral lines asymmetry has been proven as an important tool for exploration of the large-scale gas motions, temperature structure of stellar atmosphere, solar and stellar convection dynamics and granulation as well \citep{atroshchenko}. The lines asymmetry studies boosted after the high precision solar spectrum measurements had been accomplished. As it was mentioned by \citet{marmolino} the spectral lines asymmetry can also be considered as velocities gradient measure, however, a series of simplifying conditions should be fulfilled in that case. Also, some difficulties with binding of the determined velocities to certain depths in the atmosphere  should not be ignored. Under the  conditions of real atmosphere the situation is far more complex, and the inverse methods do not require such simplifications to reproduce altitude dependencies of the parameters being investigated.

This work presents the results of studying the field of velocities (both vertical and microturbulent ones) in the superadiabatic layer and overshooting convection region of the solar photosphere at small scales. One can use the obtained results to test theoretical models of stellar convection. The work is connected to the problem of convection influence on the physical conditions in solar atmosphere as well as to the problem of radiative transfer in the inhomogeneous medium.

The field of velocities exploration has been performed in the framework of the real solar granulation using profiles with high spatial resolution and considering the NLTE effects. Tikhonov's stabilizers were included into the inverse code to improve the reliability of the reproduced data.

\section{Problem definition}

The reconstruction of the parameters of inhomogeneous atmosphere  was carried out by \citet{stod02}, who solved the inverse radiative transfer problem with use of modified response functions.
In order to damp the spurious oscillations of solutions and make them insensitive to the initial estimations of parameters we used proper Tikhonov's stabilizers \citep{press}. These stabilizers improved the reliability of obtained results  \citep{stodilka}. In this case the merit function is 
$$
\chi^{2}=\chi^{2}_{0}+ \alpha S \eqno(1)
$$
where  $\chi^{2}_{0}$ is a standard merit function \citep{ruiz}, which is a measure of the experimental and theoretical line profiles closeness, $S$ is Tikhonov's stabilizer (or linear combination of stabilizers), $\alpha$ is a regularization parameter.

Tikhonov's stabilizers allow to get smooth solutions and to include prior information concerning the dependence under consideration, also they substantially improve the convergence of the NLTE inverse problem \citep{stodilka}. The linearization of (1) gives correction to the parameters, which describe intermediate inhomogeneous atmosphere model 
$$
\delta \textbf{x} = -\frac{\nabla x{_0^2}(\textbf{x}_0) + 2\alpha H \textbf{x}_0}{D(\textbf{x}_0) + 2\alpha H},
$$
where $\textbf{x}_0$ is an initial estimation of the model parameters, $H$ is matrix analogous to the stabilizer, $D$ is Hessian matrix. This model was refined iteratively.
	
To evaluate the total pressure we solved the hydrodynamic equilibrium equation under condition  of the horizontal balance  of the total pressure on the bottom boundary. In our approach the gas pressure stratification is recalculated for every variation of the temperature or field of velocities.
	
The non-equilibrium multilevel radiative transfer problem was solved  by means of the accelerated $\Lambda$-iteration method, also the solution convergence was accelerated and  multiple nodes for each intermediate model were used. The inelastic collisions with neutral hydrogen atoms were included in the statistic equilibrium equation \citep{steenbock}. The obtained NLTE parameters were used in the inverse problem.
	
Because of the limited spatial resolution of the observations it is impossible to reproduce field of velocities at small scales. Hence, such a classical parameter as microturbulent velocity has been used, which is varying with the altitude and allows for the lines broadening. Also, we used velocity of matter motion along the line of sight $(V_{los})$, which is mainly determined by convective and undulating motions. This parameter evokes additional line widening and forms its asymmetry. Macroturbulent velocity, which allows for the line broadening by macromotions was not taken into account. In our case macroscopic motions are represented by $V_{los}$. 

\section{Observational data}

Here we use the results of observations of N. Shchukina on the 70-cm German Vacuum Tower Telescope (VTT) placed on the Canary Islands \citep{kostik, khomenko}. Spectral line of neutral Iron with $\lambda=523.42$ nm has been chosen for observations; the region of its formation (by emission contribution functions) extends  from several kilometers up to 500 km by height. The observations were taken around the centre of solar disk in non-perturbed region. The image tremor on the input slit of the spectrograph did not exceed $0^{''}.5$ during the observations, i. e. spatial resolution equals to 350 km. The images were corrected for the dark current and inhomogeneous sensitivity of the pixels.

The average line profile was obtained by averaging of spectral tracks followed by co-ordination with Liege Atlas. That allows determination of true continuum level and makes binding to the corresponding wavelengths \citep{khomenko}. Thus, the set of profiles normalized on the average continuum was obtained. The set consists of 256 profiles in total corresponding to the extent of 64000 km over the surface of the Sun. The inverse procedure was applied for each profile reproducting the stratification of temperature and field of velocities $(V_{micro}, V_{los})$  in solar photosphere along two spatial coordinates: its depth $h$ and $X$, the coordinate along the spectrograph slit.

\section{Results}

The neutral Iron line profiles with $\lambda=523.42$ nm enable us to derive the distribution of the solar atmosphere parameters along the $X$ spatial coordinate at the heights from -50 km up to 500 km. Beyond this range of altitudes the response functions of this line are weakly sensitive to the variations of the parameters studied.

\begin{figure}
 \includegraphics[width=84mm]{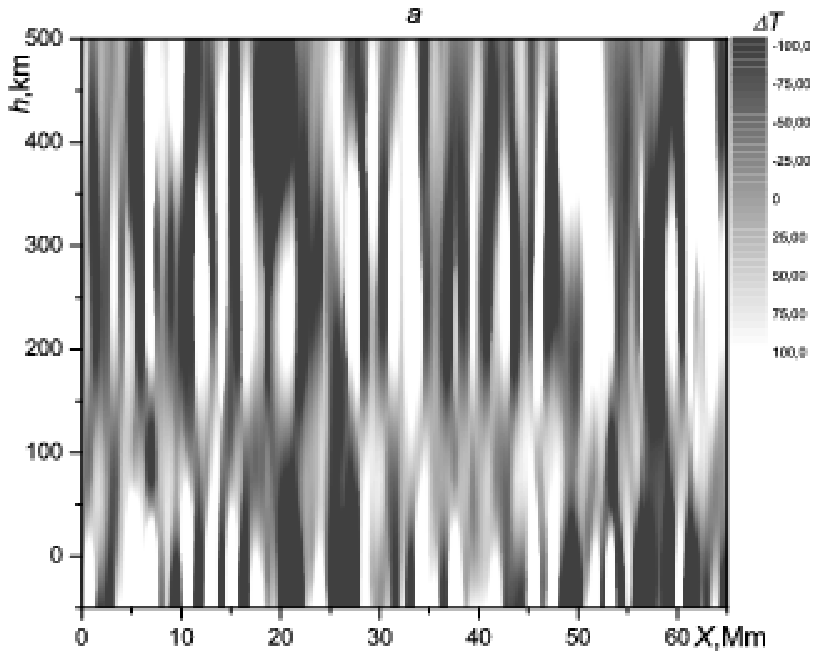}
 \includegraphics[width=84mm]{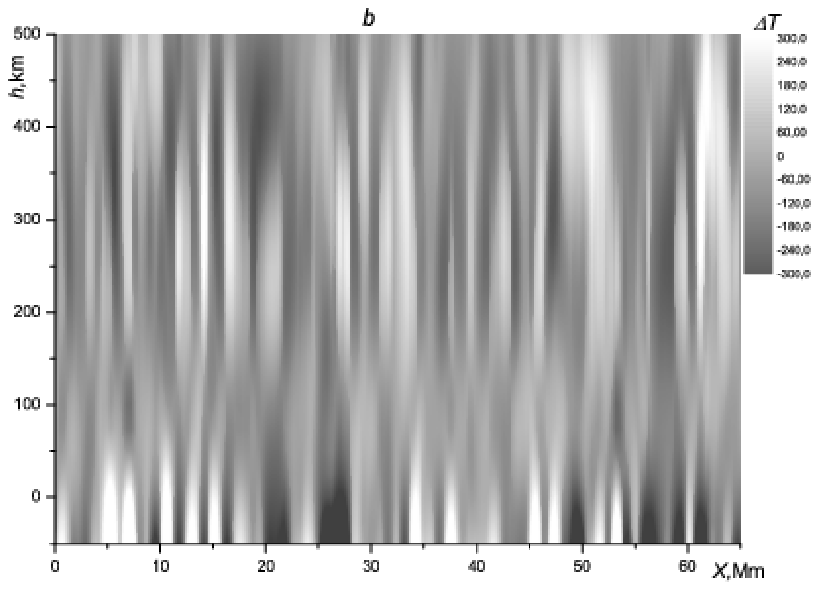}
 \caption{Distribution of temperature variations in solar granulation: a) high contrast; b) low contrast. The range of $\Delta T$ variations is constrained by $\pm 100$ K and $\pm 300$ K in order to gain better contrast in the upper layers of atmosphere. The same refers to the spatial distribution of velocities.}
  \label{one}
\end{figure}

The reconstructed distributions of temperature fluctuations, microturbulent velocity and vertical component of the solar plasma velocity are depicted in the Figures~\ref{one}, ~\ref{two}, and ~\ref{three} correspondingly.
According to Fig.~\ref{one}, where light shade depicts positive and dark one depicts negative temperature fluctuations, the horizontal temperature fluctuations take the largest values in the lower photosphere. 

Initially, those fluctuations diminish with the altitude to the value of about $\pm 100$ K at the altitude of $h\approx 100$ km and then start growing. The beginning of the overshooting convection region corresponds to the layers at $h\approx 100\div 120$ km where the first temperature inversion occurs. At this point matter of the convection cells becomes colder than in the intergranular lanes, although it continues to move upward. The same behaviour is observed for intergranules. Ascending flows are cooled down by both radiative losses and gas expansion; descending flows are heated by compression and horizontal effects \citep{gadunploner, gadun}. The temperature fluctuations in the overshooting convection region undergo intensive smoothing by radiation, especially in lower layers where the fluctuations are minimal. According to \citet{rodriguez} the temperature inversion starts at the altitude of $h\approx140$ km.

The region of ineffective convection (superadiabatic layer) is located in the lower photosphere at $h<0$ km. The superadiabaticity value $(\bigtriangledown - \bigtriangledown_{ad})$ is positive and of the order of unit, so the thermodynamic parameters fluctuations are maximal there. The Schwarzschild criterium for the upper boundary of convection zone  $(\bigtriangledown - \bigtriangledown_{ad}=0)$ is satisfied at the altitudes of $h = 0 \pm 25$ km. This criterium gives understated (for 100 km) position of the overshooting convection region relatively to the upper layer of solar convection. Classical Schwarzschild criterium does not include the influence of turbulence, taking into account the latter shifts the convective boundary upward \citep{li, robinson}.

In the upper photosphere, according to the results of solar convection simulations given in \citep{gadunploner, gadun, ploner} the role of oscillations grows. As it follows from Fig.~\ref{one} the fluctuations of smaller scales  do appear in the upper photosphere layers, i. e. besides of the overshooting convection, the oscillating motions arise in the upper layers.

\subsection{Microturbulent velocity}

\begin{figure}
 \includegraphics[width=84mm]{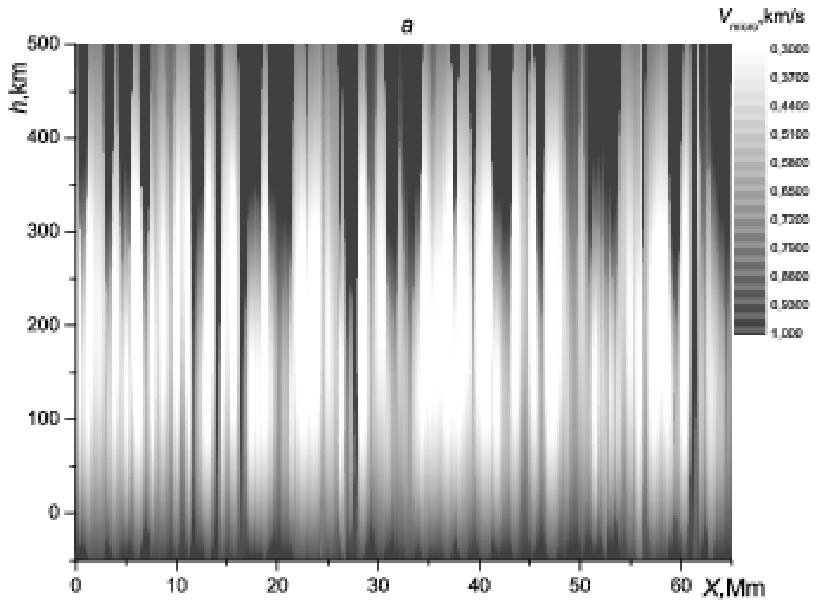}
 \includegraphics[width=84mm]{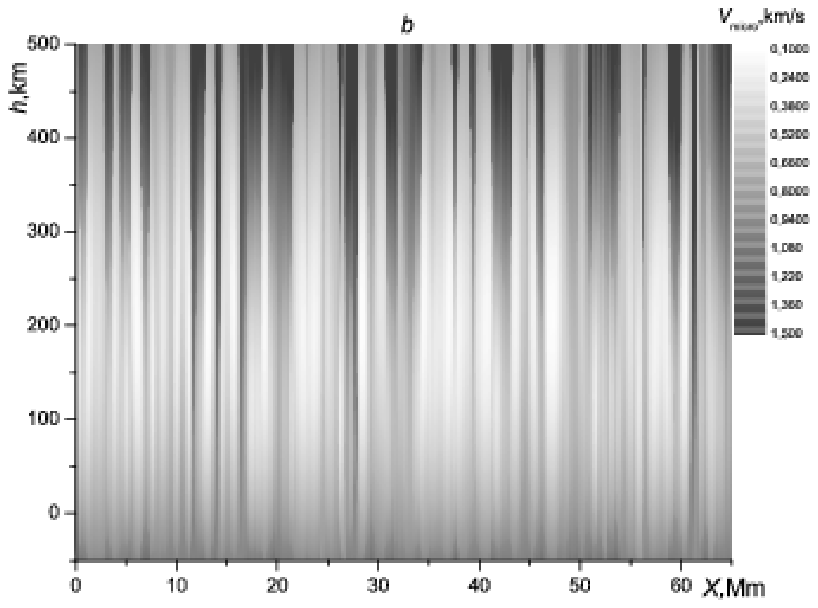}
 \caption{Microturbulent velocity distribution in solar granulation: a) high contrast; b) low contrast.}
  \label{two}
\end{figure}

2D distribution of the microturbulent velocity is presented in Fig.~\ref{two}. The microturbulent velocity in the lower and middle photosphere quite closely reflects the temperature distribution in the granular structure: $V_{micro}$ in granules is substantially lower in comparison to the intergranules. 
The microturbulent velocity in the lower photosphere takes the value about 1 km/s, below in the superadiabatic layer it is evidently larger. The beginning of the overshooting convection region is characterized by noticeable lowering of the microturbulence in the ascending flows; the microturbulent velocity drops down toward upper layers and grows up to 0.9 $\div$ 1.9 km/s value at $h>200$ km.

It is known that the decline of temperature within the region of spectral line formation leads to reducing of the equivalent line width and half-width. However, according to the observations \citep{puschmann} the half-width of spectral lines   $\lambda=6494.994$ \AA  and  $\lambda=6496.472$ \AA  FeI for large structures with angular size ranging from $1^{''}.4$ to $4^{''}.0$ (where the turbulence is developed) rises from the granule's boundary toward the centre of intergranule. Thus, an opposite tendency is observed: the rise of a line half-width corresponds to the reduction of equivalent width in the intergranules. That might be caused, on the authors' opinion \citep{puschmann}  either by enhancement of the turbulence or by presence of large velocity gradients. According to \citet{nesis93, nesis99}, enhanced turbulence in the lower photosphere concentrates on the granule boundaries while in the upper layers it spreads over the whole intergranular lane. \citet{solanki} attributed the rise of the lines half-width in the intergranules to the vertical velocities structure. Our results on the reconstruction of the field of microturbulent velocities also exhibit the enhancement of turbulence  in the intergranule, moreover, in the upper layers $(h>300$ km) and in the superadiabatic layer the region of the enhanced turbulence even expands (see Fig.~\ref{two}). 

According to the reconstruction the location of maximum turbulence in the intergranules depends on the cell's size and state. Thus, the turbulence is maximal in the central part of an intergranule if its size exceeds $1{''}.5$ and the cell undergoes an intermediate phase of its evolution, i. e. the temperature inversion occurs in the overshooting convection region and descending flow spreads over the larger part of the photosphere (by the altitude). High turbulence regions can be found also at the edges of wide cells. In the case of descending cells of smaller sizes, vertical matter flows are suppressed and a temperature contrast is low. The small-sized cells reside mainly at the initial or final stage of this development. For these intergranules the maximal turbulence is found predominantly at the edge of a cell on the interface of two flows. Thus, the high turbulence arises either at the boundary of an incipient flow drifting into the central part of the flow during its evolution or in the central part of the flow. 

In the upper photosphere where horizontal flows deviate downward to descending flows the turbulence boosts sharply and turbulent flows widen. Also they widen in the lower photosphere: increasingly dense surrounding plasma is involved by the turbulent flows' downward motion, slowing down the descending flow (see also Fig.~5). High-turbulent regions of the descending flows  narrow in lower and middle photosphere ($0 < h < 300$ km). The energy released in the motion of the descending flows feeds the turbulence \citep{nordlund}. In wide intergranules of the size $\approx 2{''}.0$ (at X = 19.6 Mm $\div$ 22.3 Mm, 25 Mm $\div$ 28 Mm, 48.5 $\div$ 50.3 Mm, 55.6 $\div$ 58 Mm) the microturbulent field of velocities suffers fragmentation into the enhanced turbulence structures related to matter flows. Apparently, microturbulent velocity is more sensitive to the fine structure of the flows than the vertical field of velocities and the temperature. Therefore, microturbulent velocity can be used for the studies of inhomogeneities as an indicator of the plasma flow structure in the solar and stellar atmospheres.

Besides of the Fraunhofer lines broadening, microturbulence plays an important role in the solar atmosphere hydrostatics. Since $P_{turb.}/P_{gas}\approx (V_{turb.}/V_T)^{2},$ where $V_{turb}=V_{micro}$ - turbulent velocity, $V_T$ - thermal velocity of matter, the relative contribution of the microturbulence into the total pressure is quite significant in the  temperature minimum region. On the other hand hydrostatic balance is rather determined by pressure gradient than by pressure itself. In the lower photosphere the gradient of $V_{micro}$ is negative to support corresponding layers, while it is positive for the upper photosphere. In that case negative gradient of gas pressure is supposed to grow up (modulo) to compensate additional down-directed force caused by the turbulence growth.

Fig.~3 depicts the temperature stratification along the vertical columns in the inhomogeneous model of the Sun; 3a) corresponds to high microturbulence (in the upper layers especially), 3b) corresponds to low microturbulence.\begin{figure}
 \includegraphics[width=84mm]{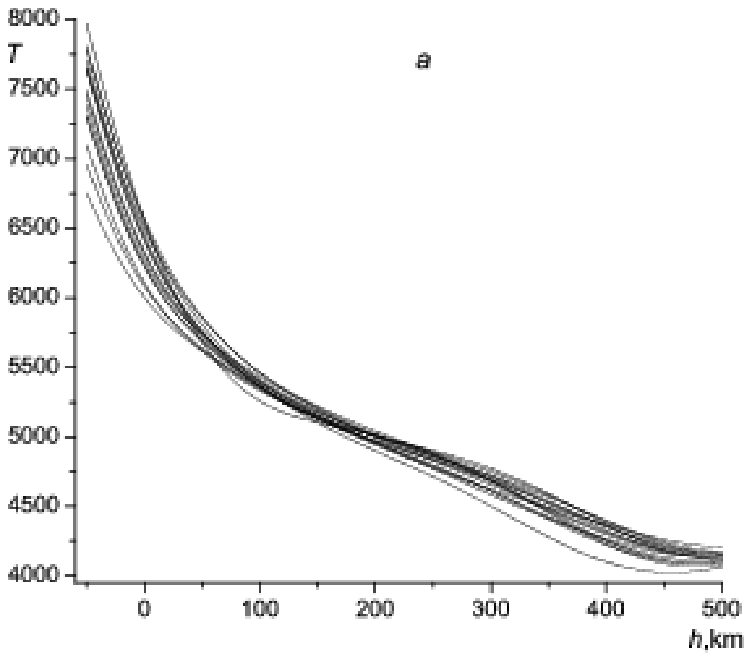}
 \includegraphics[width=84mm]{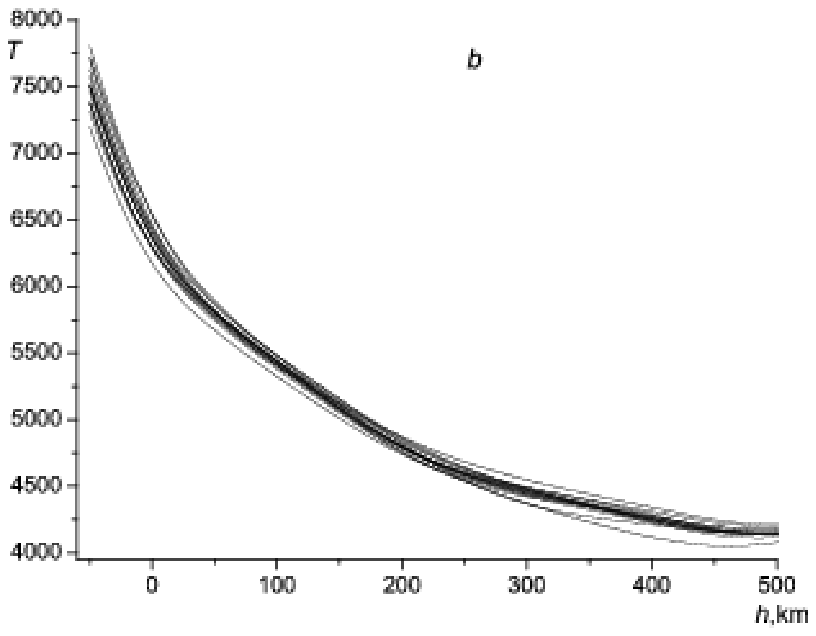}
 \caption{Altitude dependence of the temperature in solar granulation: a) high microturbulence; b) low microturbulence. }
  \label{three}
\end{figure} 
The granules with high convective velocities in the upper photosphere (see Fig.~\ref{three}a - upper part of plots at $h<100$ km) feature the enhancement of microturbulence in these layers. The temperature of these flows in upper layers drops off due to the work done by expanding gas, besides this sharp fall in temperature compensates the effects caused by the microturbulence enhancement. 

The decline of the temperature is inherent for the descending highly turbulent flows (Fig.~\ref{three}a - lower part of dependencies for $h<100$ km) causing the second temperature inversion of intergranules. The first inversion starts at the beginning of overshooting convection region (intergranules are hotter than granules), the second occurs in upper layers where the temperature of some intergranules drops sharply and they again become colder than surroundings. In the case of low microturbulence there is no sharp fall in temperature in upper photosphere (Fig.~\ref{three}b). Moreover, the major part of the columns shown in Fig.~\ref{three}b belongs to the granules.

As it follows from our results the microturbulence substantially affects the temperature structure of the inhomogeneous solar atmosphere.

We considered isotropic microturbulence model in the centre of solar disc. Anisotropic model could be obtained in the similar way by studying the edge of the disk. However, the solution of inverse problem is substantially complicated by necessity to consider not just one but several vertical columns of the inhomogeneous stellar atmosphere at the same time.

\subsection{Vertical velocity field}

When determining the absolute values of the vertical velocities it is necessary to take into account the shift of the observed lines relatively to the laboratory wavelength values or low spatial resolution lines (the line centres are known with the accuracy of $\sim 0.15$ km/s \citep{borrero}). It is also necessary to consider red-shift of Fraunhofer lines. To overcome these obstacles we corrected the velocity  assuming that the total mass flow equals to zero:
$$
\int\limits_{S}\rho V_{z}ds=0,
$$
where $S$ is the surface observed, $\rho$ is matter density.

Because the temperature inversion occurs in the overshooting convection region, some difficulties arise if flows are studied using the temperature field alone. The information about velocity field allows to determine whether two different regions, for instance with high temperature in lower photosphere and low temperature in upper photosphere or vice versa belong to the same flow or not.

\begin{figure}
 \includegraphics[width=84mm]{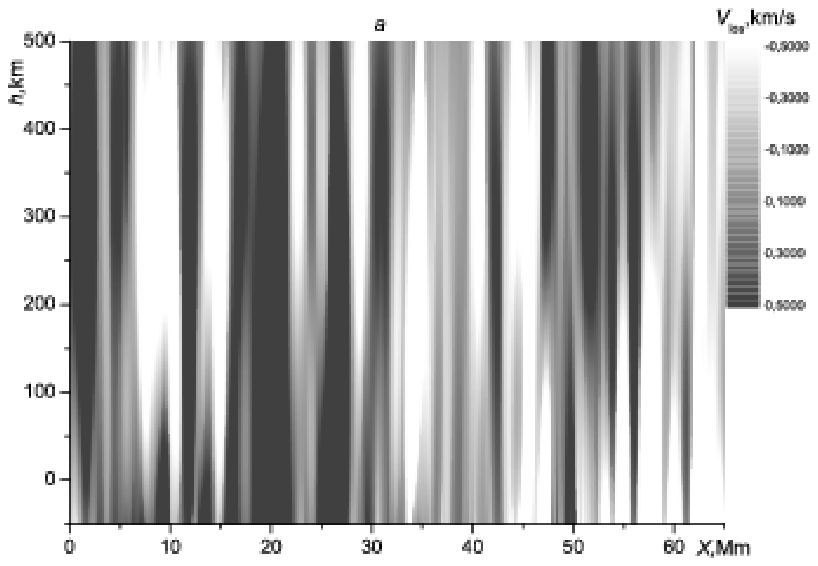}
 \includegraphics[width=84mm]{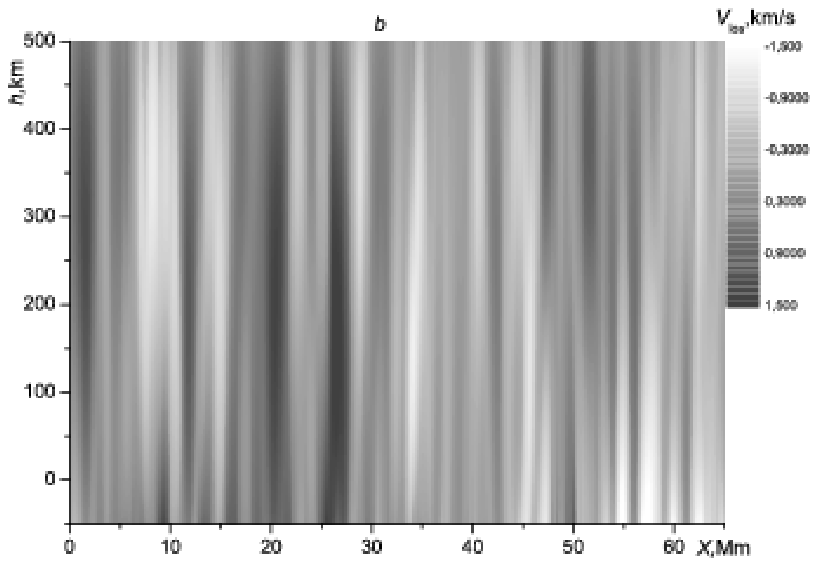}
 \caption{Vertical velocity distribution in solar granulation: a) high contrast; b) low contrast.}
  \label{four}
\end{figure}

The reproduced vertical velocity field is depicted in Fig.~\ref{four}. The light colour corresponds to the ascending flows velocity, and dark one to that of the descending flows. 
Most likely the set of flows of finite size is generated in the altitudes range considered. Following the reproduced results of the mean granulation \citep{borrero} the highly understated position of the upper limit of overshooting convection region ($h\approx255$ km) is obtained. The velocities of both convective components at the mentioned altitude are close to zero and the sign of convective velocity changes to the opposite in the upper layers, that should not be interpreted on the authors' opinion as true granulation feature. Nevertheless, in our study with high spatial resolution the convective velocity inversion  was revealed for some flows. Thus, the descending flow resides at X = 9 Mm in the lower photosphere along with ascending flows, apart from where they merge together in the upper photosphere. At X = 47 Mm in lower photosphere hot matter moves up, while in upper photosphere relatively cooler matter moves down toward the lower flow. Apparently, the granule at the beginning of its decay can be found there. The plasma, which resides on the top of the granule for a long time cools down radiatively and acquires descending velocity. As a result, the relatively small inhomogeneity of cold matter forms and starts to move downwards dragging gas from surroundings and originating in such a way the new descending flow, which is most likely 'laminar' on its initial stage (the eddies have not been formed yet). Indeed, according to the obtained results the microturbulent velocity value in that region is about 0.3 km/s, i. e. the same as in the ascending flows. Completely  formed intergranule is situated in the upper photosphere at X = 52 Mm. Its temperature is higher than that for surrounding due to the compression effects; the matter moves downward, and there is hotter matter located in the same column in the lower photosphere moving with low velocity. This is actually decayed granule remnant.

The majority of the cells scales to about 600 km by altitude, while horizontal extension of the flows amounts to $1000\div 3000$ km.

\begin{figure}
 \includegraphics[width=84mm]{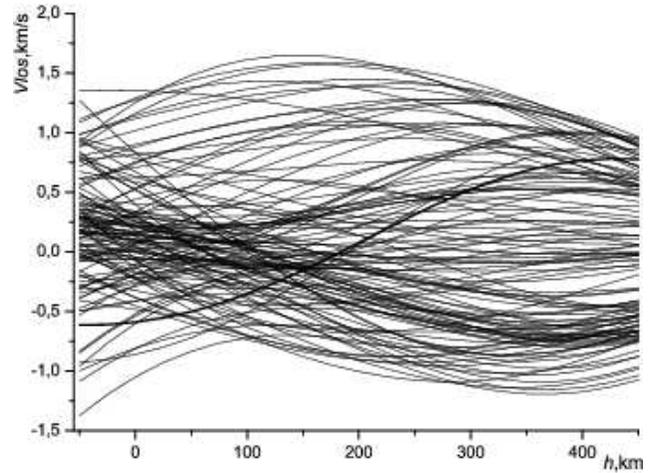}
 \caption{The altitude stratification of the vertical velocity.}
  \label{five}
\end{figure}

Altitude dependence of the cells' vertical velocities is depicted in Fig.~\ref{five}. Each line corresponds to the altitude velocity dependence for separate vertical column of the atmosphere model. 
Negative velocities correspond to the ascending flows, positive ones to the descending flows. Velocities stratification is smooth enough due to the use of stabilizers. Velocities vary in the limits of $\pm 1.0$ km/s. There is no apparent asymmetry of the vertical velocities between ascending and descending flows in the photosphere. As it can be seen from Fig.~\ref{five} there are some columns (granules) with standard altitude dependence of the velocity, i. e. the matter moves up and velocity decreases with the altitude (for the intergranules the situation is opposite), although in our case the descending flows velocities in lower photosphere are lesser than that given by \citet{frutiger}. Also, there are columns with non-standard behaviour or velocities inversion: in the lower photosphere the matter moves up while in upper photosphere it moves down (or vice versa). As it was mentioned above this is connected with the presence of cold matter (intergranule) in the lower photosphere, which moves downwards and confluences with the neighbouring ascending flows above the intergranule or with granules' decay. Besides, the gas flows are not strictly vertical because of shearing motion resulting the ingress of neighbouring flow into the vertical column. The bold line in Fig.~\ref{five} depicts the velocity distribution in the granule's centre. In the upper part of the granule the descending flow is formed due to radiative gas cooling with following formation of new intergranule. 

Significant gain of the vertical velocities amplitude fluctuations in the upper photosphere \citep{gadun} is not mentioned in our work. According to \citet{gadun}, an accurate description of radiative effects in the simulations leads to decrease of the fluctuations of temperature and vertical velocities amplitude in the photosphere, because the structure of surface layers of the atmosphere is sensitive to non-local radiative transfer.

\begin{figure}
 \includegraphics[width=84mm]{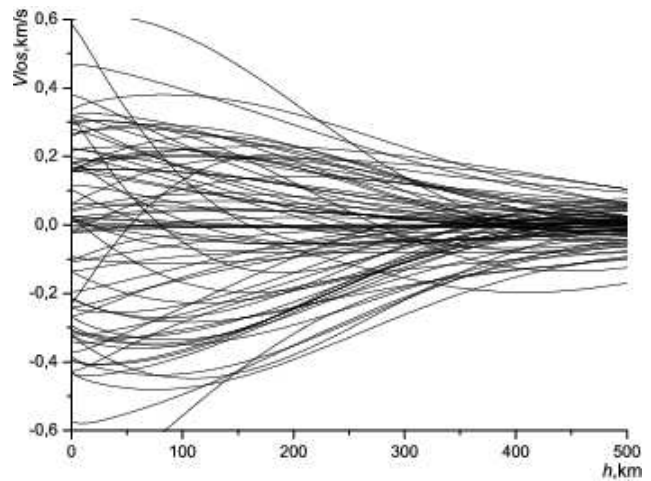}
 \caption{The altitude stratification of vertical velocities of the convective flows.}
  \label{six}
\end{figure}

Figure~\ref{six} depicts velocities of purely convective motions extracted by $k-\omega$ filtration. When moving through lower photosphere granules acquire the highest velocities due to the maximal buoyancy in superadiabatic layer, namely those layers are observed in continuum spectrum. Deceleration of the cells with altitude is typical for the ascending flows in upper photosphere and lower chromosphere. Indeed, middle and upper photosphere compound into an overshooting convection region with temperature inversion of flows, buoyancy of the cell in that region substantially decreases. \citet{rodriguez} also reveal the convective velocities decrease with altitude. The structure of convective flows described as velocity field remains steady down to the temperature minimum. Part of the convective flows reaches the temperature minimum layers.

Velocity of descending flows, which are generated in upper photosphere, increases toward deeper layers (see Fig.~\ref{six}), however there is a trend to deceleration in the lower photosphere, what apparently is connected with the compression buildup and flows' mass increment together with matter penetration into dense layers \citep{PRVBH05}.

The obtained spatial distribution of the temperature and velocity does not represent a real structure of the solar atmosphere and is rather weighted average of the phisical quantities. The smearing of the original image is produced mainly by atmospheric {''}seeing {''}. According to \citet{MS69}, the spatial smearing can be approximated by the Gaussian function; so the smeared quantity $V{'}$ is related to the unsmeared quantity $V$ as:  

$$  
V'(x_{0}) =  1/(\sqrt{2\pi}\sigma) \int\limits_{X}V(x)exp(-(x-x_{0})^{2}/2\sigma^{2})dx
$$

The value of $\sigma$ is related to measured autocorrelation length: $l \approx 1.63 \sigma$.

Thus, having the reconstructed smeared snapshots of the solar granulation one can easily obtain (e.g. by iterations) realistic spatial distribution of various physical parameters. 

Fig. 7a depicts unsmeared spatial disribution of the flows velosity obtained with $\sigma$ = 200 km; Fig. 7b depicts pixelization of the part of the same image. As it can be seen, unsmeared images possess substantionally higher contrast. Hence, at $\sigma$ = 200 km the amplitude of velocity variations increases up to 30 \%, and at $\sigma$ = 300 km even up to 50 \%. Analogously spatial smearing influences to the reproduced temperature variations.

\begin{figure}
 \includegraphics[width=84mm]{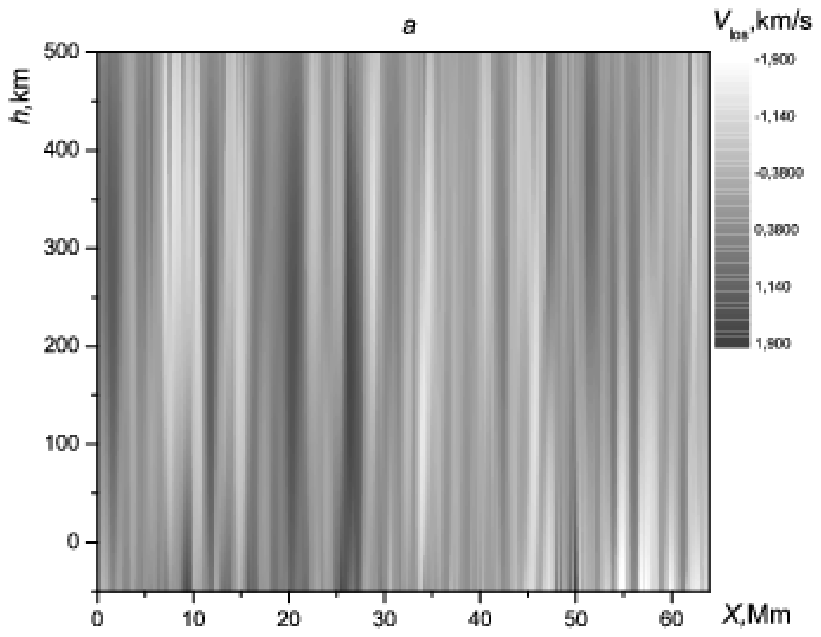}
 \includegraphics[width=84mm]{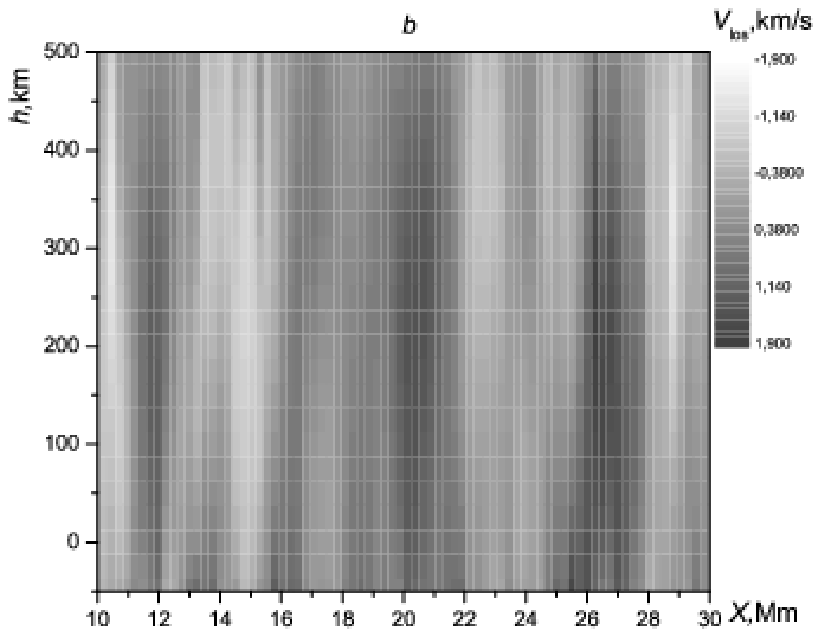}
 \caption{The effect of atmospheric seeing on solar granulation: a)unsmeared velocity snapshot; b) pixelization of the part of the same image presented in a).}
  \label{seven}
\end{figure} 

Thus, unsmeared images qualitatively coincide with smeared ones, but taking into account of  the smearing allows to study solar granulation on smaller spatial scales. 
	
Unfortunately, it is impossible to observe the lower chromosphere layers, because the profile of the line considered is not sensitive to the parameter variations in those layers.
	
Our results are obtained from the profiles with high spatial resolution while for averaged granulation studies such fine effects could not be considered. Also, let us briefly indicate some possible reasons of quantitative discrepancies between our results and the results  obtained recently by \cite{PRVBH05}. First of all, we did not use the NLTE parameters for quiet Sun model, but calculated them for every vertical column. Also, we used the hydrodynamic equilibrium equation for the total pressure instead of hydrostatic equation for gas pressure; geometrical heights scaling was done according to a requirement of horizontal balance of total pressure at bottom boundary, not only gas one. The stratification of microturbulent velocity is reconstructed simultaneously with $T$ and $V_{los}$. And last, Tikhonov's stabilizers used in our calculations improve reliability and provide higher spatial resolution.
	
\section{Conclusions}

We discussed the field of velocities features in both superadiabatic layer and overshooting convection region of solar photosphere. We realize that there is a lack of arguments for some conclusions regarding the granules motion, long-duration observations of the granulation are required for that purpose.

The field of velocities $(V_{micro}, V_{los})$ of the real solar granulation has been studied along with vertical solar photosphere structure and the granules penetration from convective zone into upper stable atmosphere layers. The results were obtained by solving the inverse NLTE problem of radiative transfer using high spatial resolution data.

On the basis of an analysis of high spatial resolution observations we have revealed following behaviour of  spatial distribution of the microturbulent velocity:
\begin{itemize}
\item The microturbulent velocity takes minimal values at the beginning of overshooting convection region and ascending  flows of matter are mostly of weak turbulence;
\item microturbulence increases sharply in the superadiabatic layer and the upper photosphere (temperature minimum region);
\item large horizontal (granule - intergranule) microturbulent velocity fluctuations are peculiarity of the upper part of overshooting convection region;
\item turbulent velocity field of wide descending flows consists of structures expanding in upper photosphere and superadiabatic layer; 
\item high-turbulent descending flows layers narrow in lower and middle photosphere ($0 < h < 300$ km);
\item maximal turbulence takes place in the central part of a cell for large intergranules at the intermediate stage of their evolution. For smaller cells maximal turbulence appears at the edge of a cell (boundary of two flows). Highly turbulent layers emerge either at the boundary of incipient flows with following drift toward the centre or in the central part of the flow;
\item broad descending flows ($> 2^{''}.0$) sustain fragmentation into the structures with increased turbulence to which the matter flows correspond;
\item the microturbulent velocity distribution can be considered as an indicator of the matter flows structure in solar atmosphere;
\item the turbulence is substantially suppressed in the ascending flows in the middle photosphere;
\item high microturbulence of the intensive flows provokes more sharp temperature decrease in upper photosphere leading to the second temperature inversion for intergranules.
\end{itemize}

It is common to the solar granulation studies to observe an inversion of vertical velocities along vertical atmosphere column being investigated. The velocity inversion can be caused by: a) the merging of the vertical flows (granules) on the disintegrated intergranule site; b) the beginning of the decay process of the granule due to strong radiative cooling of the matter in upper layers; c) shearing motions.

We revealed following behaviour of the vertical velocities in solar granulation: a) the descending flows velocity decreases in the lower photosphere; b) the convective flows velocities decrease in the middle and upper photosphere; c) part of the convective flows reaches temperature minimum layers.

The reproduced images are spatially smeared by atmospheric seeing, so real spacial variations in the solar granulation are of higher contrast. We estimated the effect of finite resolution on the spatial variations of the velocities in solar photosphere.

Finally, note that these conclusions have been derived from the reconstruction of solar granulation parameters, not by indirect statistical methods. 

One can use the obtained results also for the testing of stellar convection simulations.

\section*{Acknowledgments}
The authors are grateful to Dr. N.G.~Shchukina for kindly providing with the results of observations and to Dr. R.I.~Kostik for data reduction.

\pagebreak

\label{lastpage}

\end{document}